\newif\ifLinksBlack 
\LinksBlackfalse

\documentclass[conference]{IEEEtran}
\IEEEoverridecommandlockouts

\usepackage{cite}
\usepackage{amsmath,amsthm,amssymb,amsfonts}
\usepackage{algorithmic}

\usepackage{subfig}
\usepackage{graphicx}

\usepackage{textcomp}
\usepackage{xcolor}

\usepackage{epstopdf}

\usepackage{setspace}

\usepackage{psfrag}

\usepackage[linesnumbered,lined,boxed,commentsnumbered, ruled]{algorithm2e} 

\usepackage{array,multirow}

\ifLinksBlack
\RequirePackage[colorlinks,allcolors=black]{hyperref}
\else
\RequirePackage[colorlinks,allcolors=blue]{hyperref}
\fi

\graphicspath {{figures/}}

\def\BibTeX{{\rm B\kern-.05em{\sc i\kern-.025em b}\kern-.08em
		T\kern-.1667em\lower.7ex\hbox{E}\kern-.125emX}}

\newcommand{\I}{\mathbf{I}}

\newcommand{\U}{\mathbf{U}}

\newcommand{\Z}{\mathbf{Z}}
\newcommand{\Wbf}{\mathbf{W}}
\newcommand{\Wbfh}{\mathbf{W}^{H}}

\newcommand{\Hbf}{\mathbf{H}}
\newcommand{\Hbfh}{\mathbf{H}^{H}}
\newcommand{\Sbf}{\mathbf{\Sigma}}

\newcommand{\h}{\mathbf{h}}
\newcommand{\w}{\mathbf{w}}

\newcommand{\y}{\mathbf{y}}
\newcommand{\x}{\mathbf{x}}

\newcommand{\hatx}{\hat{\mathbf{x}}}

\newcommand{\Nscprb}{N_{\text{cs}}}

\newcommand{\fB}{f_{\text{B}}}

\newcommand{\Ap}{A_{\text{p}}}
\newcommand{\Mp}{M_{\text{p}}}
\newcommand{\Np}{N_{\text{p}}}

\newcommand{\CformRMF}{\frac{K \Mp f_{\text{B}}}{\Nscprb \Ap}}
\newcommand{\CformIIC}{\frac{f_{\text{B}}(30K^{3} + bK^2 +cK) }{N_{\text{cs}} A_{\text{p}}}}
\newcommand{\Cfilt}{\frac{\Np \Mp \fB}{\Ap}}

\newcommand{\Rpanel}{\frac{2 w_{\text{filt}} \Np \fB}{\Ap}}
\newcommand{\Rlocal}{\frac{2 w_{\text{W}} K^{2} f_{\text{B}}}{N_{\text{cs}}\Ap}}

\begin{document}
	
	\title{Processing Distribution and Architecture Tradeoff for Large Intelligent Surface Implementation}
	
	\author{
		\IEEEauthorblockN{
			Jes\'{u}s Rodr\'{i}guez S\'{a}nchez~\href{https://orcid.org/0000-0002-5531-1071}{\includegraphics[scale=0.04]{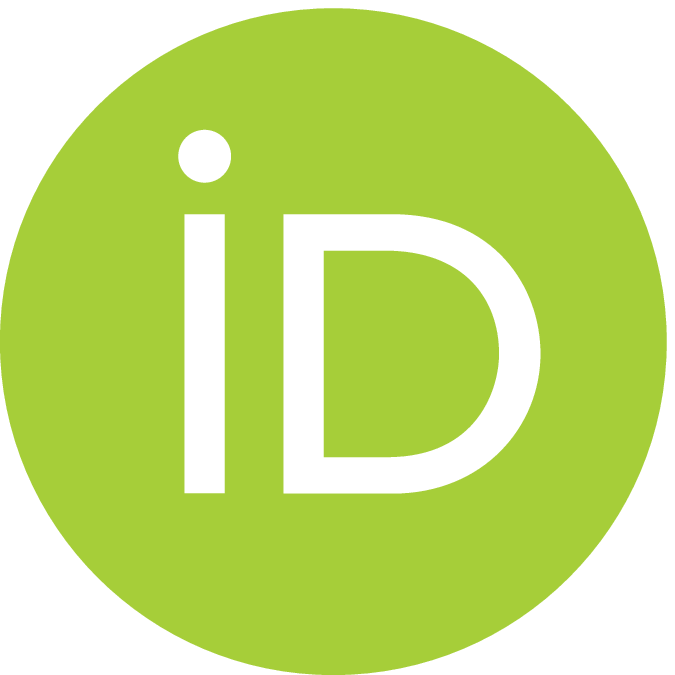}},
			Ove Edfors~\href{https://orcid.org/0000-0001-5966-8468}{\includegraphics[scale=0.04]{orcid.eps}},
			Fredrik Rusek~\href{https://orcid.org/0000-0002-2077-3858}{\includegraphics[scale=0.04]{orcid.eps}},
			and Liang Liu~\href{https://orcid.org/0000-0001-9491-8821}{\includegraphics[scale=0.04]{orcid.eps}}}\\
		\IEEEauthorblockA{Department of Electrical and Information Technology, Lund University, Sweden}
		\IEEEauthorblockA{\{jesus.rodriguez, ove.edfors, fredrik.rusek, and liang.liu\}@eit.lth.se}
	}
	
	\maketitle
	
	\begin{abstract}
		The Large Intelligent Surface (LIS) concept has emerged recently as a new paradigm for wireless communication, remote sensing and positioning. It consists of a continuous radiating surface placed relatively close to the users, which is able to communicate with users by independent transmission and reception (replacing base stations).
		Despite of its potential, there are a lot of challenges from an implementation point of view, with the interconnection data-rate and computational complexity being the most relevant. Distributed processing techniques and hierarchical architectures are expected to play a vital role addressing this while ensuring scalability. In this paper we perform algorithm-architecture codesign and analyze the hardware requirements and architecture trade-offs for a discrete LIS to perform uplink detection. By doing this, we expect to give concrete case studies and guidelines for efficient implementation of LIS systems.
	\end{abstract}

	\section{Introduction}
	\label{section:intro}
\begin{figure}
		\centering
		\includegraphics[width=0.85\linewidth]{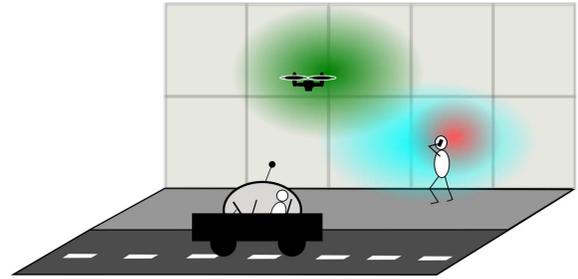}
		\caption{A LIS serving multiple users simultaneously.}
		\label{fig:LIS_concept}
	\end{figure}

	The LIS concept has the potential to revolutionize wireless communication, wireless charging and remote sensing \cite{husha_data,husha_data2,husha_asign,husha_pos} by the use of man-made surfaces electromagnetically active. In Fig. \ref{fig:LIS_concept} we show the concept of a LIS serving three users simultaneously. A LIS consists of a continuous radiating surface placed relatively close to the users. Each part of the surface is able to independently receive and transmit electromagnetic (EM) waves with a certain control, so the EM waves can be focused in 3D space with high resolution, creating a new world of possibilities for power-efficient communication.
	
	Apart from LIS, other network architectures have been proposed recently for beyond-5G systems. Some of them can be classified within the smart radio environment paradigm \cite{direnzo}, by which the wireless channel can be controlled to facilitate the transmission of information, as opposite to traditional communication systems where the channel is assumed to be imposed by nature, and transmitter and receiver adapt to changes in it. One example of this new trend is the reconfigurable surfaces, known as \textit{intelligent reflecting surfaces} (IRS), \textit{programmable metasurfaces}, \textit{reconfigurable intelligent surfaces}, and \textit{passive intelligent mirrors} among others \footnote[1]{We refer to \cite{basar} and \cite{huang_hol} for a complete list of surfaces.}, which consist of electronically passive surfaces with the capability to control how the waves are reflected when hitting their surface. Furthermore, the term LIS has also been used for such a passive surfaces \cite{taha,han,jung,huang}, with the subsequent risk of confusion. In the common form of these surfaces there is a lack of a receiver chain, therefore not having the possibility to obtain channel state information (CSI) necessary to control the reflected waves for coherence beamforming. This means that the control must come from an external system resulting in a corresponding latency. This is in conflict with the real-time requirements of many communication systems, such as cellular communications, where channel updates are required within typically 1ms. In addition, it is known that conventional MIMO communication is more efficient than IRS-aided transmission in terms of rate \cite{bjornson}. These two limitations lead us to consider LIS as the preferred architecture for beyond-5G systems.

	Regarding LIS, there are important challenges from an implementation point of view. It is known  \cite{husha_data} that a continuous LIS can be replaced by a discrete one with no practical difference in achieved capacity, and therefore making LIS implementable. This discrete LIS is made up of a large number of antennas with the corresponding receiver (and transmitter) chains producing a huge amount of baseband data that needs to be routed to the Central Digital Signal Processor (CDSP) through the backplane network. As an example, a $2m\times20m$ LIS contains $\sim 28,500$ antennas in the 4GHz band (assuming spacing of half wavelength), with the corresponding radio frequency (RF) and analog-to-digital converter (ADC) blocks. Then, if each ADC uses 8bits per I and Q, that makes a total baseband data-rate of 45.5Tbps. This is orders of magnitude higher than the massive MIMO counterpart, where this issue has been analyzed \cite{cavallaro,puglielli,jesus_journal_MaMi,muris}. In order to ensure feasibility of LIS without compromising the expected benefit over Massive MIMO, in terms of spectral efficiency (mainly due to the greater number of elements and proximity to users) there are two approaches: relax the requirements (antenna density, ADC resolution, hardware quality, etc), and design proper algorithms/architecture allowing modularization and scalability. In this paper we focus on the second approach.

	LIS is fundamentally different to massive MIMO due to the potential very large physical size of the surface and the amount of data to be handled, which requires specific processing, resources and performance analysis. \cite{juan_VTC19} is a preliminary work addressing this issue by employing a distributed approach, where panels exchange messages with neighbors in order to build the equalizers. Multiple iterations are expected to be needed until a certain level of convergence is being achieved. The lack of a need of central processing unit (while building the equalizer) in this proposal is the key argument to ensure scalability. Together with the architecture, \cite{juan_VTC19} presents the corresponding performance analysis. However, an evaluation of the required cost, from hardware point of view, is missing. For the best of our knowledge, there is not publication which performs analysis of the processing distribution, performance and the corresponding cost together for LIS.

	In this paper, we propose to tackle those challenges leveraging algorithm and architecture co-design. At the algorithm level, we explore the unique features of LIS (e.g., very large aperture) to develop uplink detection algorithms that enable the processing being performed locally and distributed over the surface. This will significantly relax the requirement for interconnection bandwidth. At the hardware architecture design level, we propose to panelize the LIS to simplify manufacturing and installation. A hierarchical interconnection topology is developed accordingly to provide efficient and flexible data exchange between panels. Based on the proposed algorithm and architecture, extensive analysis has been performed to enable trade-offs between system capacity, interconnection bandwidth, computational complexity, and processing latency. This will provide high-level design guidelines for the real implementation of LIS systems.    
	
	\section{Large Intelligent Surfaces}
	\label{section:LIS}	
	In this article we consider a LIS for communication purpose only.
	Due to the large aperture of the LIS, the users are generally located in the near field. A consequence of this is that the LIS can harvest up to 50\% of the transmitted user's power. This is one of the fundamental differences to the current 5G massive MIMO. 
	One consequence of this difference, is that the transmitted power in uplink/downlink is much lower than in traditional systems, opening the door for extensive use of low-cost and low-power analog components.
	
	Another important characteristic of LIS is that users are not seen by the entire surface as shown in Fig. \ref{fig:LIS_concept}, which can be exploited by the use of localized digital signal processing, demanding an  uniform distribution of computational resources and reduced inter-connection bandwidth, without significantly sacrificing the system capacity.	
	
	\subsection{System Model}
	We consider the transmission from $K$ single antenna users to a LIS with a total area $A$, containing $M$ antenna elements. We assume the antennas are distributed evenly with a distance of half wavelength.
	The $M\times 1$ received vector at the LIS is given by
	\begin{equation}
	\mathbf{y} = \sqrt{\rho}\Hbf\mathbf{x}+\mathbf{n},
	\end{equation}
	where $\mathbf{x}$ is the $K\times 1$ user data vector, $\mathbf{H}$ is the $M \times K$ normalized channel matrix such that $\|\Hbf\|^{2}=MK$, $\rho$ the $\mathrm{SNR}$ and $\mathbf{n} \sim \mathcal{CN}(0,\I)$ is a $M \times 1$ noise vector.
	
	Assuming the location of user $k$ is $(x_{k},y_{k},z_{k})$, where the LIS is in $z=0$. The channel between this user and a LIS antenna at location $(x,y,0)$ is given by the complex value \cite{husha_data}
	\begin{equation}
	h_{k}(x,y)=\frac{\sqrt{z_{k}}}{2\sqrt{\pi} d_{k}^{3/2}}\exp{\left( -\frac{2\pi j d_{k}}{\lambda} \right)},
	\label{eq:channel}
	\end{equation}
	where $d_{k}=\sqrt{z_{k}^{2}+(x_{k}-x)^2+(y_{k}-y)}$ is the distance between the user and the antenna, and Line of Sight (LOS) between them is assumed. $\lambda$ is the wavelength.
	
	\subsection{Panelized Implementation of LIS}
	
	\begin{figure}
		\centering
		\psfrag{panel}{Panel}
		\psfrag{ant}{Antenna}
		\psfrag{CDSP}{\tiny\color{white}CDSP}
		\psfrag{LDSP}{LDSP}
		\psfrag{PSU}{\tiny\color{white}PSU}
		\psfrag{BB}{To backbone}
		\includegraphics[width=0.7\linewidth]{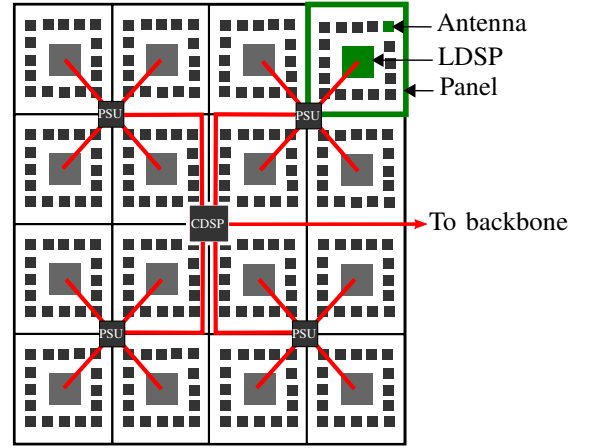}
		\caption{Overview of the LIS processing distribution and backplane interconnection. Backplane interconnection in red.}
		\label{fig:LIS_backplane}
	\end{figure}
	
	An overview of the processing distribution and interconnection in a LIS is shown in Fig. \ref{fig:LIS_backplane}. As it can be seen, we propose that a LIS can be divided into units which are connected with backplane interconnections. We will use the term $\emph{panel}$ to refer to each of these units. Each panel contains a certain number of antennas (and transceiver chains). A processing unit, named Local Digital Signal Processor (LDSP) is in charge of the baseband signal processing of a panel. LDSPs are connected via backplane interconnection network to a Central DSP (CDSP), which is linked to the backbone network. In the backplane network, there are Processing Swiching Units (PSU) performing data aggregation, distribution, and processing at different levels.     
	
	Based on the general LIS implementation framework, the number of panels $P$, the panel area $\Ap$, the number of antennas per panel $\Mp$, the algorithms to be executed in LDSP and CDSP, and the backplane topology are important design parameters we would like to investigate in this paper.
	
	\section{Uplink Detection Algorithms}
	\label{section:algo}
	
	The LIS performs a linear filtering
	\begin{equation}
	\hatx= \Wbf \y = \sqrt{\rho}\Wbf \Hbf \mathbf{x} + \Wbf \mathbf{n}
	\end{equation}
	of the incoming signal to the panels, where $\Wbf$ is the $K \times M$ equalization-filter matrix, and $\hatx$ the estimated value of $\x$.
	
	In this section we introduce two algorithms for uplink detection suitable for the panelized implementation presented in the previous section. The outcome of both is the formulation of the equalizer matrices $\{\Wbf_i\}$ for panels.

	\subsection{Reduced Matched Filter (RMF)}
	
	The Reduced Matched Filter \cite{IIC_ArXiv} is a reduced complexity version of the full MF, where the $N_{p}$ \textit{strongest} received users ($N_{p} \leq K$) by the $i$-th panel according to their respective CSI are used as filtering matrix, this is
	\begin{equation}
	\Wbf_{\text{RMF},i} = \left[ \h_{k_1}, \h_{k_2}, ...,  \h_{k_{Np}} \right]^H,
	\label{eq:W_RMF}
	\end{equation}
	where $\Wbf_{\text{RMF},i}$ is the $\Np \times \Mp$ filtering matrix of the $i$-th panel, and $\h_{n}$ is the $\Mp \times 1$ channel vector for the $n$-th user, $\{k_{i}\}$ represents the set of indexes relative to the $N_{\text{p}}$ strongest users. The corresponding strength of user $n$ is defined as $\|\h_{n}\|^{2}$
	
	\subsection{Iterative Interference Cancellation (IIC)}
	
	IIC is an algorithm that allows panels to exchange information in order to cancel inter-user interference. The detailed description of the algorithm can be found in \cite{IIC_ArXiv}, and the pseudocode for the processing at the $i$-th panel is shown below,
	\IncMargin{1em}
	\begin{algorithm}[ht]
		\SetKwInOut{Input}{Input}
		\SetKwInOut{Output}{Output}
		\SetKwInOut{Preprocessing}{Preprocessing}
		\SetKwInOut{Init}{Init}
		\Input{$\Hbf_{i}, \Z_{i-1}$}
		$[\U_z,\Sbf_z] = \text{svd} (\Z_{i-1})$\\
		$\Hbf_{eq}=\Hbf_{i} \U_{z} \Sbf_{z}^{-1/2}$\\
		$\U_{eq} = \text{svd} (\Hbf_{eq})$\\
		$\Wbfh_i=\U_{eq}(1:\Np)$\\
		$\Z_i = \Z_i + \Hbfh_i \Wbfh_i \Wbf_i \Hbf_{i}$
		\caption{IIC algorithm steps for $i$-th panel}
		\label{algo:IIC}
		\Output{$\Wbf_i, \Z_i$}
		
	\end{algorithm}\DecMargin{1em}
	where $\Hbf_i$ is the $\Mp \times K$ local CSI matrix as seen by the $i$-th panel, $\Z_{i-1}$  is the $K \times K$ matrix received from the $(i-1)$-th panel (neighbor), and $\Wbf_i$ the local filtering matrix. $\U_z$ and $\Sbf_z$ are the left unitary matrix and singular values of $\Z_{i-1}$ respectively. $\U_{eq}$ is the left unitary matrix of $\Hbf_{eq}$, and $\Wbf_i$ is made by the eigenvectors associated to the  $\Np$ strongest singular values. Each iteration of the algorithm is performed in a different panel. Matrix $\mathbf{Z}$ is passed from one panel to another by dedicated links.
	
	Ideally we would like to find the set of filtering matrices $\{\Wbf_i\}$ providing the maximum sum-rate capacity for a given channel information set $\{\Hbf_i\}$. Solving this optimization problem in a distributed way is not trivial, so in the IIC approach we solve a local optimization problem in each panel and share the result with neighbor panels. Panel $i$ will calculate $\Wbf_i$ while taking the other matrices in $\{\Wbf_i\}$ as given (fixed and not subject to optimization) in the form of $\Z_{i-1}$. This matrix $\Z_{i-1}$ acts as a noise covariance matrix in the local sum-rate optimization problem carried out locally.
	
	\section{Local DSP and Hierarchical Interconnection}
	\label{section:arch}
	In this session, we describe the corresponding LDSP and backplane architecture that supports both the RMF and IIC algorithms. We assume the OFDM-based 5G New Radio (NR) frame structure and consider uplink detection only.	

	\begin{figure*}[ht]
		\subfloat[LDSP architecture and hiarachical backplane interconnection.]{
			\includegraphics[width=0.5\linewidth]{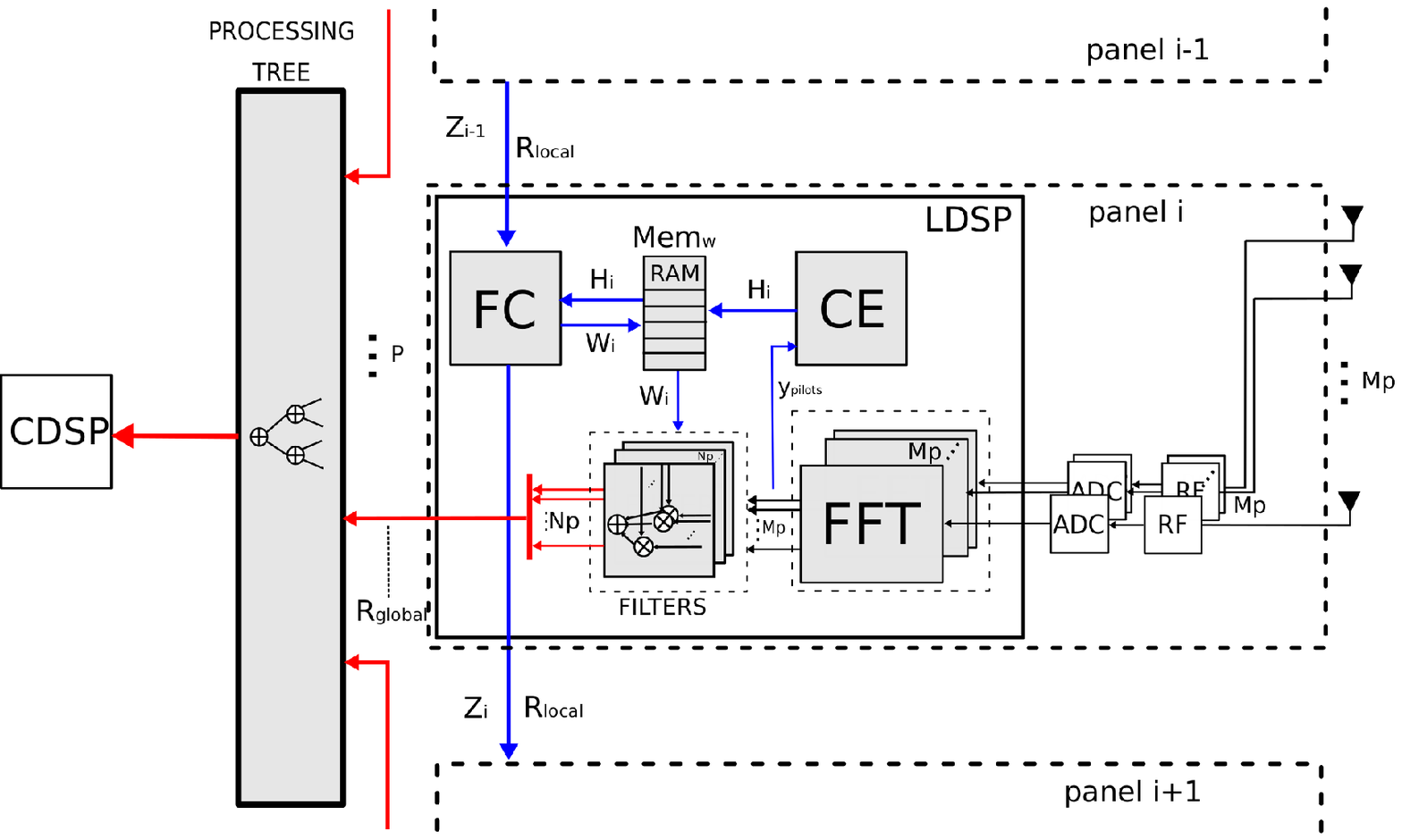}
			\label{fig:panel}
		}\hfill
		\subfloat[Tree-based global interconnection with distributed processing-switching units.]{
			\includegraphics[width=0.4\linewidth]{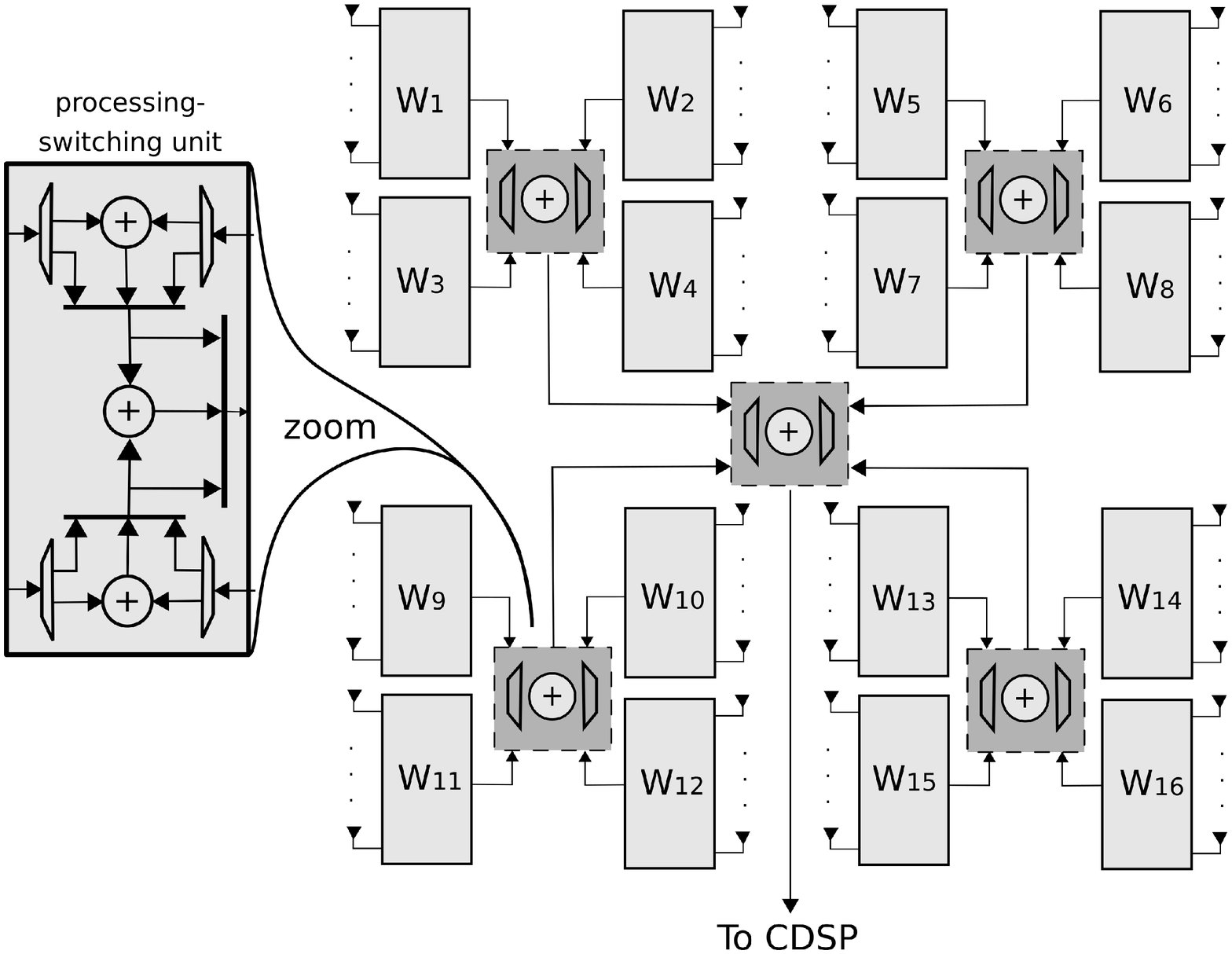}
			\label{fig:LIS_topology}
		}\\
		
		\caption{Overview of the local DSP unit in each panel and the backplane interconnection topology.}
		\label{fig:architecture}
	\end{figure*}
	
	\subsection{Local DSP in each Panel}
	The architecture of the LDSP is depicted in Fig. \ref{fig:panel}. After the RF and ADC, FFT blocks perform time-to-frequency domain transformation. The processing of the uplink signal is divided in two phases: formulation and filtering. During the formulation phase, the Channel Estimation block (CE) estimates a new $\Hbf_i$ for each channel coherence interval. In this paper we assume perfect channel estimation. The Filter Coefficient calculation (FC) block receives $\Hbf_i$ and computes the filtering matrix $\Wbf_i$. FC performs complex conjugate transpose in the case of RMF and executes Algorithm \ref{algo:IIC} in the case of IIC. $\Wbf_i$ is then written to the memory.
	During the filtering phase, the Filters block reads $\Wbf_i$ and apply it to the incoming data. The Filters block reduces the $\Mp \times 1$ input to a $\Np \times 1$ output ($\Np \ll \Mp$), which is sent to the backplane for further processing.
	
	\subsection{Hierarchical Backplane Interconnection}
	To reduced the required interconnection bandwidth, a hierarchical backplane topology is developed to fully explore the data locality in the proposed algorithms. As shown in Fig. \ref{fig:panel}, the backplane is divided into local direct panel-to-panel link (marked in blue) and global interconnection (marked in red and will be described in detail in the next sub-section). The local link is dedicated for low-latency data exchange between two neighboring panels, e.g., the $\mathbf{Z}_{i-1}$ in the IIC algorithm. The global interconnection will aggregate the $\Np \times 1$ filtering result from each panel to CDSP for final decision.
	
	\subsection{Tree-based Global Interconnection and Processing}	
	For the global interconnection, we propose to use a tree topology with distributed processing to minimize latency (the latency grows logarithmically with the number of panels), as shown in Fig. \ref{fig:LIS_topology}. There are several levels of processing switching units (PSU) in the tree to aggregate and/or combine the panel outputs. These hierarchical PSUs can reduce the overall bandwidth requirement of the backplane and also the processing load of CDSP.
	Fig. \ref{fig:LIS_topology} also shows the detailed block diagram of a PSU. It is flexible to support both RMF and IIC, and can be extended for other algorithms. Combination and bypass functionalities are used in RMF, while for IIC the streams are bypassed to the CDSP for final decision.

	\section{Implementation Cost and Simulation Results}
	\label{section:analysis}
	In this section, we analyze the implementation cost of the proposed uplink detection algorithms with the corresponding implementation architecture, in terms of computational complexity, interconnection bandwidth, and processing latency. The trade-offs between system capacity and implementation cost is then presented to give high-level design guidelines. For convenience, we summarize the system parameters in Table \ref{table:table_params}.
	
	\begin{table}[t!]
		\centering
		\begin{tabular}{l|l} 
			\hline
			$\textbf{Parameter}$ & $\textbf{Definition}$ \\
			\hline\hline
			$M_{\text{p}}$ & number of antennas per panel \\
			\hline
			$A_{\text{p}}$ & panel area \\
			\hline
			$N_{\text{p}}$ & number of filtered outputs per panel \\
			\hline
			$w_{\text{filt}}$ & bit-width of the panel output\\
			\hline
			$K$ & number of users \\
			\hline
			$f_{\text{B}}$ & signal bandwidth (Hz)\\
			\hline
			$N_{cs}$ & number of coherent subcarriers\\
			\hline
		\end{tabular}
		\caption{System parameters}
		\label{table:table_params}
	\end{table}

	\subsection{Computational Complexity}
	In Table \ref{table:C}, we summarize the required computational complexity for both RMF and ICC algorithms. The complexity includes both formulation phase and filtering phase and are normalized to panel area $A_P$. In the filtering phase, the operations are the same for RMF and ICC, which is applying a liner filter of size $N_P\times M_P$ to the $M_P\times 1$ input vector.
	
	\begin{table}[t!]
		\centering
		\begin{tabular}{c|c|c} 
			\hline
			$\textbf{Method}$ & RMF & IIC \\
			\hline
			\hline
			\multirow{2}{*}{$C_{\text{filt}}$} & \multirow{2}{*}{$\Cfilt$} & \multirow{2}{*}{$\Cfilt$} \\ 
			&& \\
			\hline
			\multirow{2}{*}{$C_{\text{form}}$} & \multirow{2}{*}{$\CformRMF$} & \multirow{2}{*}{$\CformIIC$} \\
			&& \\
			\hline
		\end{tabular}
		\caption{Computational complexity in $\text{MAC}/s/m^2$.}
		\label{table:C}
	\end{table}
	
	The formulation phase of RMF includes the computation of $\|\h\|^{2}$ for each user. For the IIC algorithm, the steps required for the formulation phase are shown in Algorithm \ref{algo:IIC}. For step 1, which consists of of a singular value decomposition (SVD) of the $K \times K$ Gramian matrix $\Z_{i-1}$, complexity is $17 K^{3}$ \cite{golub}. Step 2 has a complexity of  $(\Mp+1)K^2$, step 3 requires a complexity of  $4 \Mp^2 K + 13 K^3$, and step 4 and 5 need $\Mp K \Np + \Np K^2$. In Table \ref{table:C}, $b=\Mp+\Np+1$ and $c=4\Mp^2 + \Mp\Np$.

	\subsection{Interconnection bandwidth}
	The normalized (to panel area) bandwidth requirement for the global interconnection can be formulated as $R_{\text{global}}=\Rpanel$ [bps/$m^2$]. The corresponding bandwidth requirement for the local panel-to-panel link is (only needed for the IIC algorithm) $R_{\text{local}}=\Rlocal$  [bps/$m^2$].
	
	\subsection{Processing Latency}
	The processing latency of the filtering phase can be formulated as $L_{filtering} = T_{\text{Filter}} + \log_{4}(P) T_{\text{PSU}}$, where $T_{\text{Filter}}$ is the time needed for performing the linear filtering and $T_{\text{PSU}}$ represents the PSU processing time as well as the PSU-to-PSU communication time.
	
	The latency of the formulation phase differs for RMF and IIC. For RMF, the formulation phase is done in parallel in all the panels. The corresponding latency $L_{\text{form,RMF}}$ depends on the computational complexity $C_{\text{form, RMF}}$, the clock frequency, and the available parallelism in the computation. On the other hand, the latency for IIC includes both computation and panel-to-panel communication. The worst case is  $L_{\text{form,IIC}} = P T_{\text{compute,IIC}} + (P-1) T_{\text{panel-panel}}$, where $T_{\text{compute, IIC}}$ is the time for computing the filter coefficient and $T_{\text{panel-panel}}$ is the transmission latency between two consecutive panels.	

	\begin{figure*}[htb]
		\centering
		\subfloat[RMF method.]{
			\includegraphics[width=0.5\linewidth]{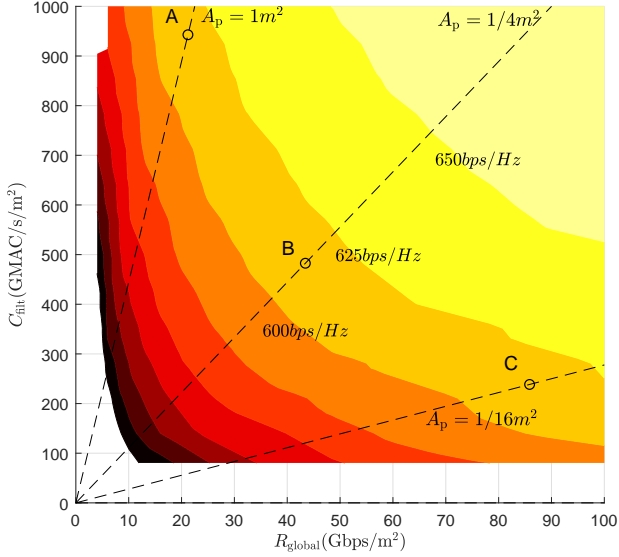}
			\label{fig:compl_vs_Rglobal_RMF}
		}
		\subfloat[IIC method.]{
			\includegraphics[width=0.5\linewidth]{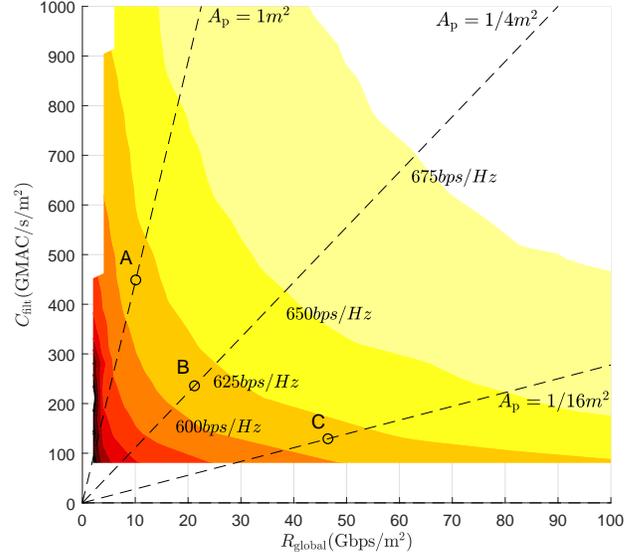}
			\label{fig:compl_vs_Rglobal_IIC}
		}
		
		\caption{Sum-rate contour plot as a function of filtering complexity ($C_{\text{filt}}$) and  inter-connection bandwidth ($R_{\text{global}}$). Carrier wavelength ($\lambda$) = $7.5cm$, number of users ($K$) = $50$, $\text{SNR} = 0dB$, signal bandwidth ($\fB$) = $100MHz$, ADC resolution ($\w_{\text{filt}}$) = $8 bits$, number of coherence subcarriers ($\Nscprb$) = $12$, and antenna spacing is $\lambda/2$.}
		\label{fig:results}
	\end{figure*}

	\subsection{Results and Trade-offs}
	\label{section:results}
	
\begin{figure}
		\centering
		\psfrag{LIS}{LIS}
		\includegraphics[scale=0.7]{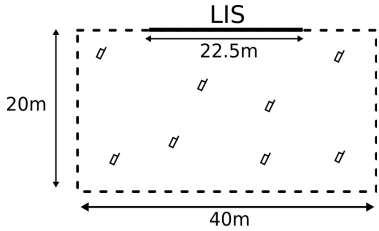}
		\caption{Top view of the simulation scenario.}
		\label{fig:LIS_scenario}
	\end{figure}

	The scenario for simulation is shown in Fig. \ref{fig:LIS_scenario}. Fifty users ($K=50$) are uniformly distributed in a $40m \times 45m$ (depth x width) area in front of a $2.25m \times 22.5m$ (height x width) LIS. Signal bandwidth and carrier frequency are 100MHz and 4GHz, respectively.

	The average sum-rate capacity at the interface between panels and processing tree for both algorithms is show in Fig. \ref{fig:results}. The figures show the trade-offs between computational complexity ($C_{\text{filt}}$ in the vertical axis) and interconnection bandwidth ($R_{\text{global}}$ in the horizontal axis). Dashed lines represent points with constant panel size $\Ap$, which is another design parameter for LIS implementation. To illustrate the trade-off, we marked points A, B, and C in the figures, presenting 3 different design choices to a targeted performance of 610bps/Hz.
	Comparing the same points in both figures, it can be observed the reduction in complexity and interconnection bandwidth of IIC compared to RMF. We can also observe as small panels (e.g., point C comparing to point A) demand lower computational complexity in expense of higher backplane bandwidth. Once $\Ap$ is fixed, the trade-off between system capacity and implementation cost (computational complexity and interconnection data-rate) can be performed depending on the application requirement.
	
	\section{Conclusions}
	\label{section:conclusions}
	In this article we have presented distributed processing algorithms and the corresponding hardware architecture for efficient implementation of large intelligent surfaces (LIS). The proposed processing structure consists of local panel processing units to compress incoming data without losing much information and hierarchical backplane network with distributed processing-switching units to support flexible and efficient data aggregation. We have systematically analyzed the system capacity and implementation cost with different design parameters and provided design guidelines for the implementation of LIS.
	
	As a future direction in our research, we aim for the implementation of a LIS, as a proof-of-concept of this technology.

	\section*{Acknowledgment}
	This work was supported by ELLIIT, the Excellence Center at $\text{Link\"{o}ping-Lund}$ in Information Technology.
	
	\nocite{ozdogan}
	\bibliographystyle{IEEEtran}
	\bibliography{IEEEabrv,LIS_architecture}
	
\end{document}